\documentclass{article}

\usepackage{arxiv}

\usepackage[utf8]{inputenc} % allow utf-8 input
\usepackage[T1]{fontenc}    % use 8-bit T1 fonts
\usepackage{hyperref}       % hyperlinks
\usepackage{url}            % simple URL typesetting
\usepackage{booktabs}       % professional-quality tables
\usepackage{amsfonts}       % blackboard math symbols
\usepackage{nicefrac}       % compact symbols for 1/2, etc.
\usepackage{microtype}      % microtypography
\usepackage{lipsum}		% Can be removed after putting your text content
\usepackage{graphicx}
\usepackage{natbib}
\usepackage{doi}

\title{Explainable Anomaly Detection for Industrial Control System Cybersecurity}

\author{{Do Thu Ha}\\
	International Research Institute for Artificial Intelligence and Data Science\\
	Dong A University, Da Nang, Vietnam\\
	\texttt{hadt@donga.edu.vn} \\
	\And
	{Nguyen Xuan Hoang} \\
	Hanoi University of Science and Technology, Vietnam\\
	\texttt{hoang.nx180085@sis.hust.edu.vn} \\
	\And
	{Nguyen Viet Hoang} \\
	Hanoi University of Science and Technology, Vietnam\\
	\texttt{hoang.nv186315@sis.hust.edu.vn} \\
	\And
	{Nguyen Huu Du} \\
	Hanoi University of Science and Technology, Vietnam\\
	\texttt{hoang.nv186315@sis.hust.edu.vn} \\	
	\And
	{Truong Thu Huong} \thanks{Corresponding author: huong.truongthu@hust.edu.vn)}\\
	Hanoi University of Science and Technology, Vietnam\\
	\texttt{huong.truongthu@hust.edu.vn} \\
	\And
	{Kim Phuc Tran} \\
	Univ. Lille, ENSAIT, ULR 2461 - GEMTEX - Génie et Matériaux Textiles, F-59000 Lille, France\\
	\texttt{kim-phuc.tran@ensait.fr} \\
}

\renewcommand{\shorttitle}{\textit{arXiv} Template}

\hypersetup{
pdftitle={Explainable Anomaly Detection for Industrial Control System Cybersecurity},
pdfsubject={},
pdfauthor={Do Thu Ha, Nguyen Xuan Hoang, Nguyen Viet Hoang, Nguyen Huu Du, Truong Thu Huong, Kim Phuc Tran},
pdfkeywords={XAI, LSTM Autoencoder,  Anomaly Detection, ICS, Gradient SHAP},
}

\begin{document}
\maketitle

\begin{abstract}
	Industrial Control Systems (ICSs) are becoming more and more important in managing the operation of many important systems in smart manufacturing, such as power stations, water supply systems, and manufacturing sites. While massive digital data can be a driving force for system performance, data security has raised serious concerns. Anomaly detection, therefore, is essential for preventing network security intrusions and system attacks. Many AI-based anomaly detection methods have been proposed and achieved high detection performance, however, are still a "black box" that is hard to be interpreted. In this study, we suggest using Explainable Artificial Intelligence to enhance the perspective and reliable results of an LSTM-based Autoencoder-OCSVM learning model for anomaly detection in ICS. We demonstrate the performance of our proposed method based on a well-known SCADA dataset.

Copyright \copyright ~ 2022, IFAC (International Federation of Automatic Control) 
\end{abstract}

% keywords can be removed

\section{Introduction}
An industrial control system (ICS) is a smart manufacturing system that controls and monitors the operation of geographically dispersed remote facilities, through sensors and intelligent computing technologies (\cite{lu2020smart}, \cite{kusiak2018smart}). Such ICSs play a vital role for the national infrastructure suppliers such as electricity, water, gas, and so more. Historically, these systems operated on proprietary hardware and/or software, physically isolated from external connections,  but they have recently adopted information technology and wireless communication to increase connectivity over the internet. Accordingly, modern technologies like the Internet of Things (IoT), Cloud Computing, and Artificial Intelligence (AI) are integrated into ICSs for enhancing system performance, but this also increases the risk of cybersecurity vulnerabilities ~\cite{sakhnini2021security}. As on-site facilities are directly controlled by ICSs, when a cyber-intrusion incident occurs, it may cause physically and economically severe repercussions. Hackers, for example, can gain control of the network and install malicious code to damage system operations. Along with various techniques applied in practice for solving network intrusion problems like Intrusion Detection System (IDS) and the firewall,  AI-based anomaly detection (AD) has also been investigated and evolved, aiming to cope with more novel and complex threats in ICSs. Anomaly detection (AD) algorithms tend to make satisfactory prediction outcomes ( \citet{zaidi2019performance, tran2016efficiency}).
AD is associated with seeking data without an expected pattern or with a pattern different from the normal stream. The AD methods brings efficiency even if a new attack or abnormal behavior arises, thus outperforming conventional classifiers. %Relying on data characteristics, AD can be approached in non-time series or time series and the references for the AD method are abundant. One can refer to \citet{DAS2020101935, luca2020} and \citet{chen2021joint} to name a few. 

Although the AI-based anomaly detection technologies demonstrate high efficiency, the model reliability could be diminished if the cause of the model-predicted results is not provided, which usually arises from its black-box problem. In practice, the explanation of such AD models is critical because a predicted anomaly does not necessarily come from an actual cyber-attack, it could come from another technical problems. A typical example is that abnormalities can derive from malfunctions of sensors (vibrations, temperature,...), where abnormal sensor readings can be indicative of impending failures. As a result, when anomalies occur, analyzing the decision-making of the AD model is time-consuming. This problem is also a significant challenge surrounding cyber-attack detection in ICSs, as most previous studies have not been concerned with the interpretability of the model’s detected results. One way to solve the problem is to use Explainable AI (XAI), a technique that helps people to understand or explain exactly the process inside an AI algorithm. The use of XAI in interpreting anomaly detection outcomes has been considered recently in the literature, see  \citet{hwang2021sfd} and \citet{morichetta2019explain}, for example.

In this study, we aim to build an XAI module for a method of AD in the context of ICSs. In particular, we consider a hybrid model which is a combination of the  LSTM Autoencoder and one-class support vector machine (OCSVM) algorithms suggested in \citet{nguyen2021forecasting}. Thanks to the separating anomalies from the data outputted by the LSTM Autoencoder network, this hybrid model has shown very good performance in detecting anomalies in sales from the fashion industry based on multivariate time series data. We, therefore, suggest using the model to detect anomalies in ICS. The performance of the method is evaluated based on a well-known dataset, namely the Gas Pipeline dataset (\cite{morris2014industrial}). The obtained result shows that the proposed model outperforms some recent other AD solutions in the same ICS context. Furthermore, to overcome the black-box issue mentioned above, an XAI module for LSTM Autorencoder - OCSVM is proposed. The use of XAI for interpreting this kind of AI algorithm for AD has not been considered before in the original work as well as most other AD proposals for ICSs in the literature. 

Based on this approach for AD, the security specialists can analyze the features contributing the most to the predicted anomaly via visual explanations. It also allows to find the real cause by suggesting physical components or sensors that are more likely to be involved in the predicted anomaly. By checking those devices, they can verify whether the anomaly comes from an actual cyberattack and make timely responses. As a result,  it is possible to save time and maintenance costs, and improve the reliability of the proposed AD model. 
The rest of the paper is organized as follows. In Section \ref{Sec: 2}, we present the proposed method for AD that aims to satisfy both conditions: providing high performance and being transparent. In particular, we describe the hybrid LSTM Autoencoder - OCSVM model and the XAI algorithm. An example to illustrate the use and the performance of the proposed method is given in Section 3. Section 4 is for some Concluding remarks and perspectives.

\section{Related works}
In \cite{das2020anomaly}, the authors presented Logical Analysis of Data (LAD), a supervised classification method, to design an AD system deployed on the Secure Water Treatment (SWaT) data. \cite{alfeo2020using} proposed a reliable deep learning-based approach that combines an autoencoder and a discriminator based on a general heuristics processing on time series data. Among many methods that are effective for detecting abnormal data, the Long Short Term Memory (LSTM)network is a prominent and very popular method. \citet{tran2019anomaly} showed that their LSTM-based approach leads to higher performance in terms of accuracy and a lower percentage of false alarm rate, compared to a traditional machine learning method. \cite{patil2019explainable} utilized LSTM architecture for sequential anomaly detection which is obtained by processing log files, and analysed the prediction results using layer-wise relevance propagation. The use of the LSTM algorithm for anomaly detection at different times in the system can also be seen in \cite{essien2020deep,liu2020deep}. Recently, LSTM has been combined with another AI algorithm to deal with time-series data, named Autoencoder, see, \cite{chen2021joint} and \citet{gjorgiev2020time}, for example.
\section{Explainable Anomaly Detection for Industrial Control Systems}
\label{Sec: 2}
In this section, we will describe the  hybrid LSTM- based Autoencoder - OCSVM model for AD as well as the XAI based module for interpreting the model black-box.
\subsection{Anomaly Detection using the hybrid LSTM- based Autoencoder - OCSVM model}
\begin{center}
\begin{figure*}[hbt]
    \centerline{\includegraphics[width=17cm]{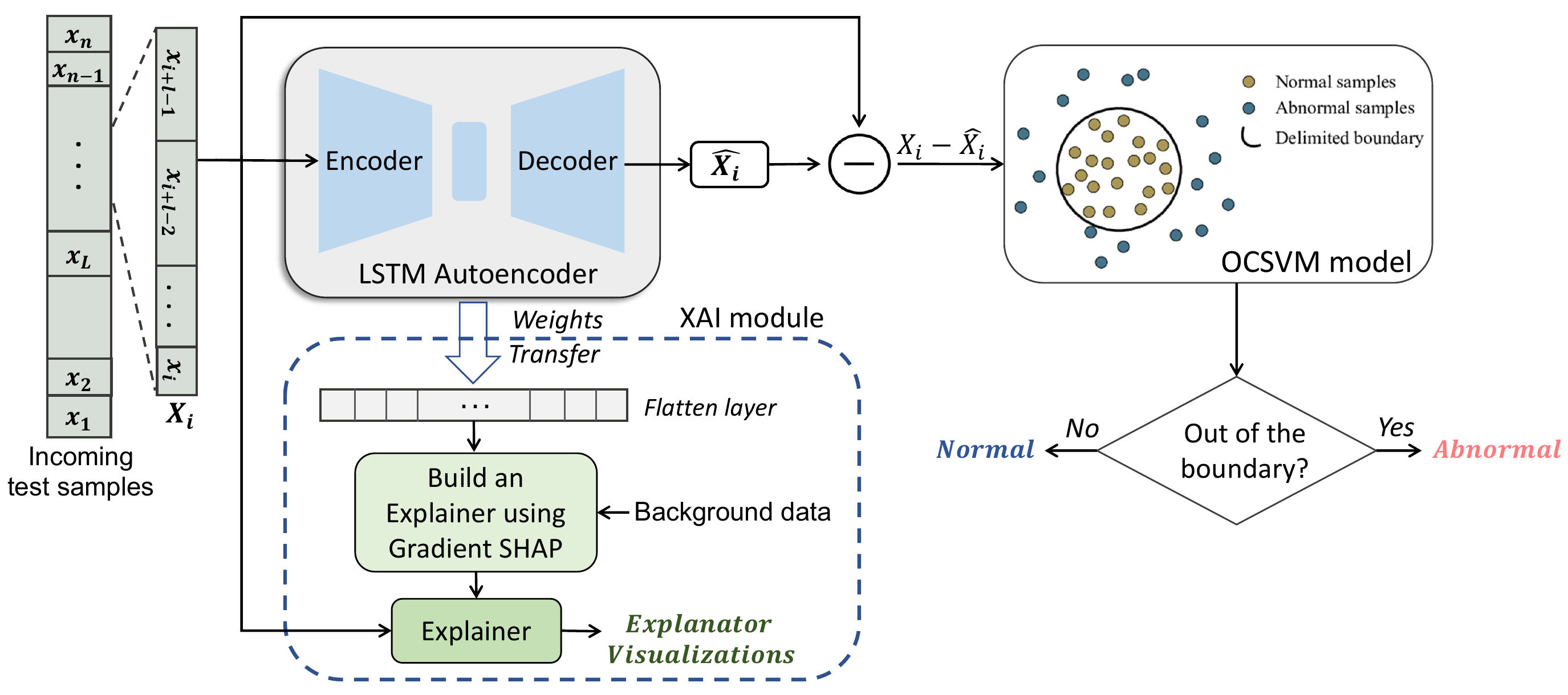}}
   \caption{An XAI-based module is proposed for interpreting the black-box anomaly detection model }
    \label{Modelling}
\end{figure*}
\end{center}

LSTM Autoencoder refers to an Autoencoder network that applies LSTM for both parts of the encoder and decoder. The objective of the autoencoder is to learn a  compressed representation for the input using encoding and decoding. More specifically, in an Autoencoder network, the input is compressed in the encoder part into the code, and then it is reconstructed by decompressing the code in the decoder part. The obtained output will be compared to the input and the error is examined and back-propagated through the network to update the weights. With the use of the LSTM cell for both the encoder and decoder, the LSTM Autoencoder benefits from both models: it outperforms the regular autoencoder in dealing with input sequences. Several perspectives of the LSTM Autoencoder for unsupervised anomaly detection in time series are discussed in \citet{provotar2019unsupervised}. The use of LSTM Autoencoder for similar topics can be seen in 
\citet{pereira2018unsupervised} and \citet{principi2019unsupervised}.

Although the LSTM Autoencoder has many advantages to deal with the time series data, there are still situations in which the algorithm is not sufficiently effective.  As mentioned above, after decoding, the output from the decoder part is compared to the input and the difference is examined. Once a big difference is recognized, i.e. the reconstruction loss is significantly high, the data point can be considered an anomaly.  In this process, it is necessary to find a threshold that allows for determining which difference is large enough. This is not a simple task and the performance of the method could be significantly affected if there is no effective way to choose this value. To overcome this difficulty,  \citet{nguyen2021forecasting}  suggested combining the LSTM Autoencoder network with the one-class support vector machine (OCSVM). By this suggestion, the LSTM Autoencoder is used only for extracting crucial information from the data, and the task of detecting the abnormal objects from the input is left to the OCSVM. OCSVM can be considered as a variant of the traditional SVM for the case we have only one class and the goal is to test a new data point to decide if it is normal.

Suppose that the Autoencoder LSTM-OCSVM has been trained from a normal sequence. During inference, denote the incoming data is $X = \{x_{1}, x_{2}, \ldots, x_{n}\}$, where $x_j \in \mathbb{R}^{m}$, $j = 1,2,\ldots,n$; and $m$ is the number of features. As illustrated in Figure \ref{Modelling}, $X$ is then divided into $n - l +1$ overlapping sequences $X_{i}=\{x_{i}, x_{i+1}, \dots, x_{i + l - 1}\}, i = 1,2,...,n-l+1$, by using a sliding window of size $l$. The window size should be chosen based on a grid search on a range of values of the power of 2 as well as the goal of finding the best performance in the trade-off with running time. The difference between $X_i$ and its reconstructed representation $\hat{X_i}$ is passed through the trained OCSVM for AD.  
 
The performance of the hybrid LSTM Autoencoder-OCSVM algorithm has been demonstrated in \citet{nguyen2021forecasting}: it works well for both simulated data and real data. That means, the algorithm can be applied to solve the problems related to anomaly detection in several fields in practice. There is only one question that should be answered in the use of this method: how does it decide if the input is anomalous. A solution for the question relying on the XAI will be discussed in the sequel.

\subsection{Integration of Explainable Artificial Intelligence}

Explainable Artificial Intelligence (XAI) refers to the algorithms that enable humans to comprehend the AI models, leading to trust in the output of the applied model and also the effective management of the benefits that the model provides. XAI is able to describe an AI model, its expected impact, and potential biases, namely, it meets our expectations for an AI model, which is functioning as expected, producing transparent explanations, and is visible in how they work.

In the literature, there are several techniques for XAI, and one of the prominent techniques is SHapley Addictive exPlanations (SHAP), a game-theory-based approach for globally or locally interpreting the output of any ML or DL model introduced in \cite{lundberg2017unified}.
This technique estimates the contribution level of each feature to the predicted results by using Shapley values, i.e., the average expected marginal contribution of a feature over all the possible combination sets or coalitions. The main idea is to clarify a complex model $f$ based on a more comprehensible approximate model $g$. Similar to LIME in \cite{ribeiro2016should}, explanation models tend to employ simplified input variables $x'$, mapping to the original samples $x$ via a function $h_x$ that satisfies $x = h_x(x')$, and local explanation methods make sure
$f(h_x(z')) \approx g(z')$ whenever $z' \approx x'$. The model $g(z')$ can be constructed as follows:
\begin{equation}
 g(z') = \phi_0 + \sum_{k=1}^{M} \phi_k z'_k,  \label{explanatory_model}
\end{equation}
where, $M$ is the number of simplified features, $z' \in \{1,0\}^M$ is a coalition containing the simplified values $z_k$ of input features, and $\phi_k \in \mathbb{R}$   implies the Shapley values of each $k^{th}$ feature. Accordingly, the value $z_k' = 1$ signifies the presence of the feature in the coalition and vice versa $z_k'= 0$. 
The value $\phi_k$ can be computed by:
\begin{equation} 
\phi_k=\sum_{\mathcal{S} \subseteq N \backslash \{k\} } \frac{|\mathcal{S}|!(|N|- |\mathcal{S}| -1)!}{|N|!}
\big( v(\mathcal{S} \cup \{k\}) -v(\mathcal{S})) \label{Shap_value}
\end{equation}
where, $N$  is the set of all features, $\mathcal{S}$ is a subset of $N$ excluding the $k^{th}$ feature, and $v(\mathcal{S})$ is the prediction value for features in the subset $\mathcal{S}$. 
For each sliding window, Shapley values $\phi_k$ in each of its samples are calculated, and the contribution of the features at the global perspective is given based on these values. 

Technically, the SHAP framework employs a variety of computing methods depending on the model type$ f$ which needs to be explained. These methods have been classified as the model-agnostic approximation methods (e.g., Linear SHAP (\citet{vstrumbelj2014explaining}), Kernel SHAP (\citet{lundberg2017unified})) and  model-type-specific  approximation methods
(e.g., Tree SHAP (\citet{lundberg2020local}), Deep SHAP (\citet{lundberg2017unified}), and Gradient SHAP). 

Among many methods of the SHAP framework, the Deep SHAP and the Gradient SHAP are computationally configured for deep learning model types, especially models taking the three-dimension data as input. While the Deep SHAP utilizes a SHAP-values-estimating algorithm,  the Gradient SHAP approximates SHAP values by computing the expectations of gradients by randomly sampling from the distribution of baselines or references.  Since the Deep SHAP is more time-consuming during processing, the Gradient SHAP will be applied in this study.

In the use of the Gradient SHAP method, the AI model output is expected to be in the form of a vector or a single value. However, the output of LSTM -Autoencoder is not the case. Therefore, in the XAI module, we build another model with the weights transferred from the original pre-trained model, while the structure is modified a bit by adding a flattened layer on top of this model, as shown in Figure \ref{Modelling}. The explainer of Gradient SHAP is formed by taking this model and a background dataset from the training set as inputs. When an anomaly is predicted, the anomalous window can be fed into this built explainer to obtain explanatory visualizations. The way the LSTM Autoencoder-OCSVM works for detecting anomalies based on a SCADA dataset will be demonstrated in the sequel. Then we apply the Gradient SHAP to comprehend the output. With the visualizations of computed Shapley values, we can discover the contribution ratio of the features to the prediction values and acquire some benefits of XAI, especially in such a gas pipeline infrastructure. 

\section{Illustrative example }
\begin{figure}[hbt]
    \centerline{\includegraphics[width=9cm]{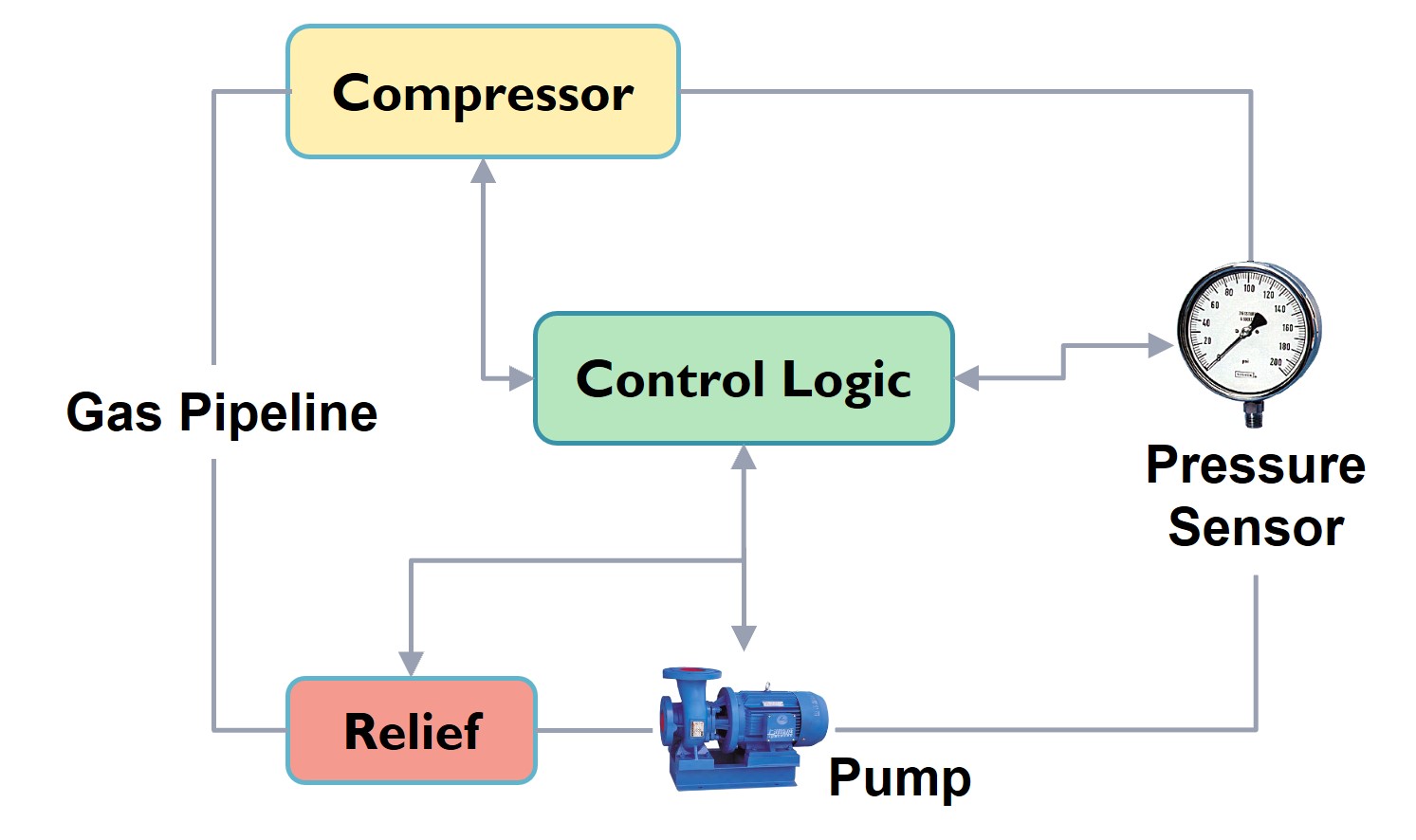}}
   \caption{Gas pipeline system diagram}
    \label{Gaspipeline}
\end{figure}

In this section, we present the performance and interpretation of the proposed approach based on a real-world dataset logging the network traffic of the SCADA system, namely, the Gas Pipeline dataset in \cite{morris2014industrial}. The components of the simulated scheme from the SCADA include a gas pipe, compressor, pressure sensor, pump, and pressure relief valve controlled by a solenoid, which has been displayed in Figure \ref{Gaspipeline}. 
The components in the system communicate by using the MODBUS protocol, and a proportional integral derivative controller (PID) helps to maintain the required pressure level. Table \ref{dataset} characterizes 17 features contained in each record, where some signify information about the network such as the MODBUS packet length and MODBUS function code, while the remainder shows the industrial process state readings measured by the sensors.

\begin{table}
\caption{Feature Descriptions in the Gas Pipeline dataset}
\label{dataset}
\begin{center}
\begin{tabular}{|p{0.35\linewidth} | p{0.55\linewidth}|}%{|l|l|}
\hline
\multicolumn{1}{|c|}{\textbf{Feature}} & \multicolumn{1}{c|}{\textbf{Description}}  \\ \hline
address  & The station address of the MODBUS slave device \\ \hline
crc rate & Packet checksum value\\ \hline
function & MODBUS fnction code  \\ \hline
length & MODBUS packet length \\ \hline
setpoint  & The pressure set point when the system is in the
Automatic system mode \\\hline
gain & PID gain \\ \hline
reset rate & PID reset rate \\\hline
deadband & PID dead band \\\hline
cycle time & PID cycle time \\\hline
rate & PID rate \\\hline
system mode & Operating mode: off\textendash{0}, manual\textendash{1}, automatic\textendash{2} \\\hline
control scheme & Pressure control: pump\textendash{0}, solenoid\textendash{1} \\\hline
pump & Pump closed\textendash{0} or opened\textendash{1} \\\hline
solenoid & Valve closed\textendash{0} or opened\textendash{1}\\\hline
pressure measurement & Pipeline pressure value \\\hline
command response & Package: response\textendash{0}, command\textendash{1} \\\hline
time & Time stamp \\\hline
\end{tabular}
\end{center}
\end{table}

In our experiment, the percentage of the training set and test set is 80\% and 20\%, respectively. Accordingly, the training data set with 40192 normal samples are fed into the model for training. The performance of the LSTM Autoencoder-OCSVM model is assessed on 10048 normal and abnormal samples, with the percentage of anomalies of roughly 20\%, which is measured by standard metrics: Precision, Recall, and F1-Score. In the context of industrial control systems,  detecting all actual anomalies tends to be more critical since any anomaly may cause it. As a result, we aim to optimize the Recall and F1-score metric rather than the Precision metric since in comparison with damage costs caused by anomalies, a trade-off with a tiny number of false alarms is acceptable. In addition,  the training time should be considered in this trade-off to meet the real-time feature of the AD model. In our experiment, the optimal sliding window size has been found using the grid search.

The obtained results from our proposed model relying on these features are then compared with the performance of another different algorithms in the literature for detecting anomalies from the same SCADA dataset, including  K-means\textendash{CAE}  ( \cite{chang2019anomaly}), the SVM  (\cite{anton2019anomaly}), the LSTM  (\cite{feng2017multi}) the NB  ( \cite{shirazi2016evaluation}), and %SAE-DNN-DT and
 the DT, DNN, and RF ( \cite{al2020ensemble}). As can be seen in Table  \ref{performance}, the LSTM Autoencoder-OCSVM model outperforms all the other methods in terms of Recall of  96.28\%. Moreover, our proposed model also achieved a significant F1-score of 90.12\% in comparison with the others. These figures demonstrate the LSTM Autoencoder OCSVM model is highly viable for industrial control system cybersecurity.
%In the context of smart manufacturing,  detecting all actual anomalies tends to be more critical because any anomaly may cause severe damage to the assets, facilities, even humans. As a result, with the Recall metric of $99.11 \% $, the LSTM Autoencoder OCSVM  model is highly feasible and promising in ICSs.\\

\begin{table}
    \centering
    \caption{The Performance of the LSTM Autoencoder - OCSVM method based on the SCADA dataset compared to other methods in the literature}
    \hspace*{0mm}
  \scalebox{0.9}{
    \begin{tabular}{|c|c|c|c|c|c|c|c|c|c|}
      \hline
\textbf{Method} &  \textbf{Precision} & \textbf{Recall} & \textbf{F-score}\\
 \hline
% \textbf{LSTM Autoencoder\textendash{OCSVM}}  & \textbf{ 0.9799} & \textbf{0.9848} & \textbf{0.9911} & \textbf{0.9635} \\
% \textbf{LSTM Autoencoder\textendash{OCSVM}}  &  \textbf{0.8048} & \textbf{0.9926} & \textbf{0.8889} \\
% LSTM Autoencoder-KQE &  0.9210 &  0.9298 & 0.9791 & 0.8410\\
\textbf{LSTM Autoencoder\textendash{OCSVM}}  &  \textbf{0.8470} & \textbf{0.9628} & \textbf{0.9012} \\
K-means\textendash{CAE} & 0.9543 & 0.8352 & 0.8908\\
SVM &  0.7820 & 0.9360 & 0.8520\\
LSTM  & 0.9400 & 0.7800 & 0.8500\\
NB  & 0.8195 & 0.7692 & 0.8595\\
% SAE-DNN-DT & 0.9600 & 0.9463 & 0.9372 & 0.9383\\
DT  & 0.9159 & 0.6808 & 0.7239\\
DNN  & 0.8994 & 0.6389 & 0.6709\\
RF  & 0.9142 & 0.6298 & 0.6591\\
\hline
  \end{tabular} }  
\label{performance}
\end{table}

\begin{figure}[hbt]
    \centerline{\includegraphics[width=9cm]{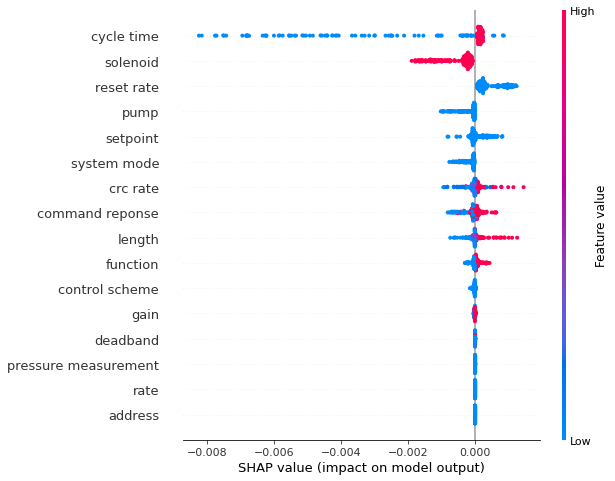}}
   \caption{The global interpretation for one sequence including abnormal patterns}
    \label{SHAP1}
\end{figure} 

As mentioned above, a good learning model should ensure both high performance and transparency. The next step in our experiment is to use Gradient SHAP to understand how the model works, i.e we analyze the causes of the detected anomalies via Gradient SHAP-based visualizations. Note that the feature 'time'  is also omitted in our data pre-processing. Figure \ref{SHAP1} demonstrates the influence of each feature in one  sequence (i.e, one sliding window) including abnormal patterns, analyzed on the model output from a global perspective. In this visualization, the features are ranked in descending importance order, the horizontal line denotes whether the impact of that feature (i.e, SHAP value) is associated with a lower or higher predicted value for that sample. Accordingly, a negative SHAP value expresses that the higher the feature value is, the lower the predicted value is, and vice versa, a positive SHAP value implies that the higher the feature value is, the higher the predicted value is.  It can be seen that, overwhelmingly,  lower  "cycle time" feature values profoundly impact the predicted anomaly, whilst this is corresponding to higher "solenoid" feature values. That is, the engineers could gain more insight into their smart manufacturing system as well as the constraint between the anomaly and features' respective physical components.

In practice, since an anomaly can arise from various threats that are not necessarily a  cyber-intrusion incident, engineers must conduct an entire system inspection to validate whether the anomaly comes from an actual cyberattack or not. This process seems to take time, energy, and even money. Nevertheless, it could be overcome based on the visualization shown in Figure \ref{SHAP1} . In this figure, the impact of  'cycle time', 'solenoid', and 'resets rate'  features are the most significant on the predicted anomaly. Combined with the descriptions shown in Table \ref{dataset}, it can be inferred that checking the behavior of the PID controller and the solenoid valve should be prioritized. Also, the pump should be then considered. During maintenance, if all of the physical components are still working properly, it can be concluded that the real cause of the predicted anomaly is an actual cyberattack. Thus, our proposal succeeds in supporting engineers or cybersecurity experts in investigating anomalies predicted by the black-box AD model. In addition to helping to save time and maintenance costs, it also fulfills the transparency of the model, which promotes engineers or operators to trust the proposed AD model. Thanks to these positive interpretations, XAI is expected to be integrated into other black-box-related AD studies for ICSs in the future. The code source utilized in this study is available
in \href{https://github.com/Nxhoang56/Explainable-Anomaly-Detection-forIndustrial-Control-System-Cybersecurity}{our Github}.

\section{Concluding remarks and perspectives}
With the booming of the IoT era, industrial control systems are increasingly evolving thanks to modern Internet communication technologies. Nonetheless, they are the top target for network security intrusions or system attacks, leading to severe implications. As a result, anomaly detection is crucial, applied to various smart production lines. In this paper, we deploy an LSTM Autoencoder-based model, combined with OCSVM to spot anomalies in industrial control systems. Our evaluation in the illustrative demonstration shows that the proposed model achieves prominent performance with the Recall metric of $96.28\%$ over the Gas Pipeline SCADA dataset. The results also indicate that our solution outperforms several existing works. Besides, we provide a visual, comprehensive explanation of the detected anomaly.  This not only increases the transparency of the black-bock model but also enables experts to have a deeper insight into the system, thereby enhancing the reliability of the anomaly detection model. Finally, it is our concern for further research to investigate other XAI methods for anomaly detection techniques in industrial control systems.

\section*{Acknowledgement}
  This work was supported by Hanoi University of Science and Technology (HUST) under Project T2021-PC-010.

\bibliographystyle{unsrtnat}
\bibliography{references}  
\end{document}